\documentclass[12pt]{article}
\usepackage{microtype} \DisableLigatures{encoding = *, family = * }
\usepackage{graphicx,multirow}
\usepackage{amsmath}
\usepackage{amsfonts,amssymb}
\usepackage{natbib}
\bibpunct{(}{)}{;}{a}{,}{,}
\usepackage[bookmarksnumbered=true, pdfauthor={Wen-Long Jin}, breaklinks=true]{hyperref}

\usepackage{hyperref} 
\hypersetup{
    colorlinks=true,
    linkcolor = blue,
    anchorcolor = blue,
    citecolor = blue,
    filecolor = blue,
    urlcolor = blue,
    pdfpagemode=FullScreen,
    }

\newcommand{\commentout}[1]{}

\newcommand{\ba}{\begin{array}}
        \newcommand{\ea}{\end{array}}
\newcommand{\bc}{\begin{center}}
        \newcommand{\ec}{\end{center}}
\newcommand{\bdm}{\begin{displaymath}}
        \newcommand{\edm}{\end{displaymath}}
\newcommand{\bds} {\begin{description}}
        \newcommand{\eds} {\end{description}}
\newcommand{\ben}{\begin{enumerate}}
        \newcommand{\een}{\end{enumerate}}
\newcommand{\beq}{\begin{equation}}
        \newcommand{\eeq}{\end{equation}}
\newcommand{\bfg} {\begin{figure}[htbp]}
        \newcommand{\efg} {\end{figure}}
\newcommand{\bi} {\begin {itemize}}
        \newcommand{\ei} {\end {itemize}}
\newcommand{\bqn}{\begin{eqnarray}}
        \newcommand{\eqn}{\end{eqnarray}}
\newcommand{\bqs}{\begin{eqnarray*}}
        \newcommand{\eqs}{\end{eqnarray*}}
\newcommand{\bsl} {\begin{slide}[8.8in,6.7in]}
        \newcommand{\esl} {\end{slide}}
\newcommand{\bsq}{\begin{subequations}}
        \newcommand{\esq}{\end{subequations}}       
\newcommand{\bss} {\begin{slide*}[9.3in,6.7in]}
        \newcommand{\ess} {\end{slide*}}
\newcommand{\btb} {\begin {table}}
        \newcommand{\etb} {\end {table}}
\newcommand{\bvb}{\begin{verbatim}}
        \newcommand{\evb}{\end{verbatim}}

\newcommand{\m}{\mbox}
\newcommand {\der}[2] {{\frac {\m {d} {#1}} {\m{d} {#2}}}}

\newcommand {\pd}[2] {{\frac {\partial {#1}} {\partial {#2}}}}

\newcommand{\cas}[1]{{{\left \{ \ba #1 \ea \right. }}}

\newcommand{\reff}[1] {{{Figure \ref {#1}}}}
\newcommand{\refe}[1] {{(\ref {#1})}}

\def\la      {{\lambda}}

\def\pmb#1{\setbox0=\hbox{$#1$}%
   \kern-.025em\copy0\kern-\wd0
   \kern.05em\copy0\kern-\wd0
   \kern-.025em\raise.0433em\box0 }

\newtheorem{theorem}{Theorem}[section]

\newtheorem{lemma}[theorem]{Lemma}

\def\la     {{\lambda}}

\usepackage[draft]{changes}
\definechangesauthor[name={Wenlong Jin},color=red]{WJ}

\usepackage{multirow}
\usepackage{subcaption}
\usepackage{float}

\begin{document}
\author{Wen-Long Jin \footnote{Department of Civil and Environmental Engineering, California Institute for Telecommunications and Information Technology, Institute of Transportation Studies, 4000 Anteater Instruction and Research Bldg, University of California, Irvine, CA 92697-3600. Tel: 949-824-1672. Fax: 949-824-8385. Email: wjin@uci.edu. Corresponding author} and Irene Martinez \footnote{Transport \& Planning, Delft University of Technology, Stevinweg 1, Delft, 2628 CN, The Netherlands. Email:  I.Martinez@tudelft.nl.}}

\title{Monotone three-dimensional surface and equivalent formulations of the generalized bathtub model} 

\maketitle

\begin{abstract}
In the Lighthill-Whitham-Richards (LWR) model for single-lane traffic, vehicle trajectories follow the first-in-first-out (FIFO) principle and can be represented by a monotone three-dimensional surface of cumulative vehicle count. In contrast, the generalized bathtub model, which describes congestion dynamics in transportation networks using relative space, typically violates the FIFO principle, making its representation more challenging.

Building on the characteristic distance ordering concept, we observe that trips in the generalized bathtub model can be ordered by their characteristic distances (remaining trip distance plus network travel distance). We define a new cumulative number of trips ahead of a trip with a given remaining distance at a time instant, showing it forms a monotone three-dimensional surface despite FIFO violations. Using the inverse function theorem, we derive equivalent formulations with different coordinates and dependent variables, including special cases for Vickrey's bathtub model and the basic bathtub model.

We demonstrate numerical methods based on these formulations and discuss trip-based approaches for discrete demand patterns. This study enhances understanding of the generalized bathtub model's properties, facilitating its application in network traffic flow modeling, congestion pricing, and transportation planning.
\end{abstract}

{\bf Keywords}: Generalized bathtub model; Relative space; Monotone three-dimensional traffic surface; Inverse function theorem; Equivalent formulations.

\section{Introduction}

Describing how vehicles and trips move in a road network is essential for urban traffic management and planning \citep{cascetta2009transportation}. In the celebrated Lighthill-Whitham-Richards (LWR) model \citep{lighthill1955lwr,richards1956lwr} for a single-lane road, vehicle trajectories are described in the traditional absolute space with respect to the traveling distances on a road, as shown in \reff{fig:absolute_relative_space}(a), where three trips (solid red, dashed blue, and dotted green) start from different origins, end at different destinations, and use different routes. The LWR model can be represented by a  three-dimensional surface in terms of the cumulative number of vehicles passing a location by a time instant \citep{makigami1971traffic}. In this model, vehicles follow the first-in-first-out (FIFO) principle and are ordered by the car-following rule. As a result, the corresponding three-dimensional surface is monotone: increasing in time and decreasing in space.
Then, from the inverse function theorem, one can derive equivalent formulations in flow, trajectory, and schedule coordinates \citep{leclercq2007lagrangian,laval2013hamilton}, which form the basis for developing more sophisticated models for human-driven and automated vehicles and managing traffic flow on a road segment \citep{jin2016equivalence}.  

\bfg\bc
\includegraphics[width=6in]{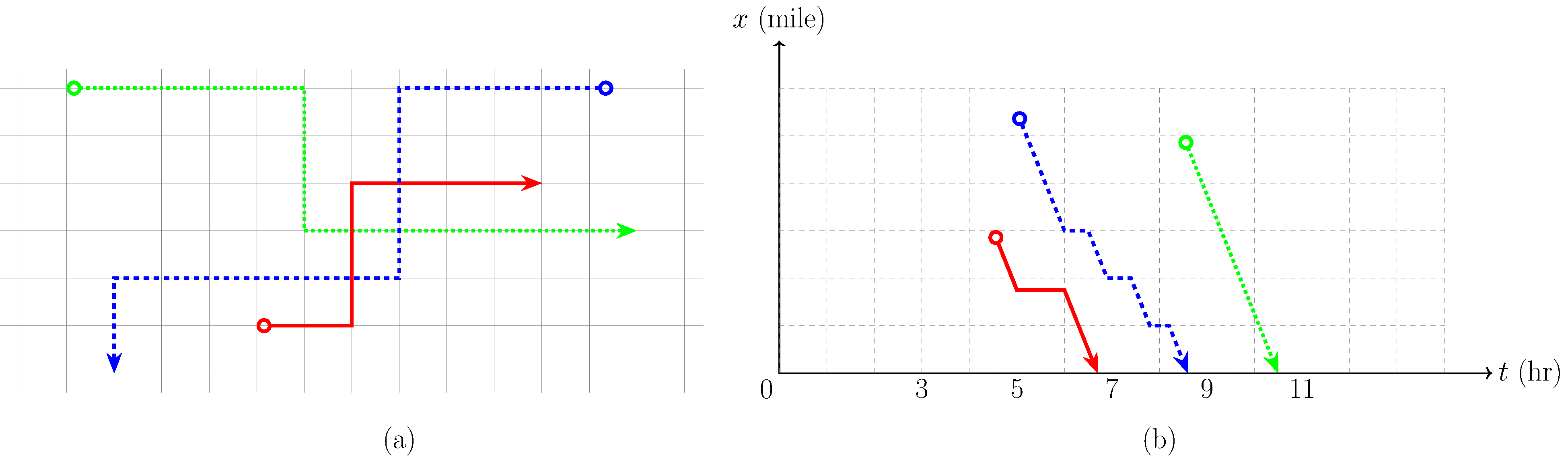}
\caption{Illustration of trips in different spaces: (a) Absolute space within a grid network; (b) Relative space based on remaining distances to individual trip destinations}\label{fig:absolute_relative_space}
\ec\efg

The LWR model has been extended to describe traffic dynamics in a road network via the Cell Transmission Model \citep{daganzo1995ctm,lebacque1996godunov} and the Link Transmission Model \citep{yperman2006mcl}. Incorporating merging and diverging rules, they are systems of partial or delay differential equations \citep{jin2017riemann,jin2015ltm}, respectively. Therefore, the calibration, computation, and analysis of these models are costly. In the past decades, a new approach emerged by viewing the whole network as a single reservoir or bathtub \citep{vickrey1991congestion, vickrey2020congestion,small2003hypercongestion,daganzo2007gridlock,arnott2016equilibrium,jin2020generalized}. A key idea is to describe all trips' trajectories in the unified relative space with respect to their remaining distances, as illustrated in \reff{fig:absolute_relative_space}(b), where the three trajectories (solid red, dashed blue, and dotted green) correspond to those in \reff{fig:absolute_relative_space}(a). Comparing the two presentations of the trips in two spaces, we can see that the absolute space tracks the origins, destinations, and routes of individual trips, but cannot easily describe the traveling speeds. In constrast, the relative space ignores the exact locations of the origins, destinations, and routes, but tracks the entering times, the exiting times, and the remaining distances to the destinations of individual trips, from which we can derive the traveling speeds at specific times. In the two representations, the trip lengths should be consistent with each other, but all other information is complementary. In particular, the relative space offers a unified approach to describing traveling speeds of individual vehicles and congestion dynamics in a road network. Therefore, the bathtub models have been applied to study departure time choice and design network-wide or cordon congestion pricing, traffic control, and other management and planning schemes \citep{vickrey2020congestion,geroliminis2013optimal,jin2021compartmental,balzer2023dynamic,martinez2024dynamic}. Refer to \citep{johari2021macroscopic} for a comprehensive review of such models and their applications.

In particular, assuming that all trips share the same travel speed and that the speed is given by a network fundamental diagram \citep{godfrey1969mechanism,geroliminis2008eus}, \citep{jin2020generalized,jin2021introduction} derived a generalized bathtub model for travel demands given by general distributions of trips' entering times and distances. This is a nonlocal partial differential equation in the cumulative number of trips at a time instant with a remaining distance not smaller than a value. It was shown that Vickrey's bathtub model \citep{vickrey1991congestion, vickrey2020congestion,small2003hypercongestion,daganzo2007gridlock} is a special case of the generalized bathtub model with a time-independent negative exponential distribution of trip distances; and that the basic bathtub model \citep{arnott2016equilibrium,arnott2018solving} is a special case where all trips have the same distance. However, different from the LWR model, the generalized bathtub usually violates the FIFO principle, and the three-dimensional surface in the cumulative number of trips at a time instant with a remaining distance not smaller than a value is not monotone in time. Thus, one cannot obtain similar equivalent formulations as those for the LWR model with the inverse function theorem.

Inspired by the monotone three-dimensional surface representation of the LWR model, and recognizing the many potential applications of the generalized bathtub model in network traffic flow modeling and control, in this article we aim to develop an analogous monotone three-dimensional surface and equivalent formulations for the generalized bathtub model despite its violation of the FIFO principle.  First, building on the previously established characteristic distance ordering from \citep{jin2020generalized}, we observe that, in the generalized bathtub model, trips in a network can be ordered with respect to their characteristic distances, which equal the remaining trip distances plus the network travel distances. Thus, we refer to those with smaller characteristic distances as being ahead of others. Even though the cumulative number of trips passing a location by a time instant on a virtual road was defined for the basic bathtub model in \citep{jin2020generalized},  the FIFO principle is satisfied in this case, and the variable is the same as that for the LWR model \citep{makigami1971traffic}. In this study, we then define a cumulative number of trips ahead of a trip with a remaining distance at a time instant for the generalized bathtub model. It can be shown that the new cumulative number leads to a monotone three-dimensional surface. Then, based on the monotone relationships between time, network travel distance, characteristic trip distance, and the cumulative number of trips, we use the inverse function theorem to derive equivalent formulations with different coordinates and dependent variables. We also present the equivalent formulations for Vickrey's bathtub model \citep{vickrey1991congestion, vickrey2020congestion}  and the basic bathtub model \citep{arnott2016equilibrium,arnott2018solving}. We further demonstrate how these equivalent formulations can be used to devise numerical methods and discuss equivalent trip-based formulations of the generalized bathtub model when the travel demand is given by discrete pairs of entering times and trip distances. 

The rest of the article is organized as follows. In Section 2, we define a new cumulative number of trips ahead of a trip with a remaining distance at a time instant and demonstrate that it leads to a monotone three-dimensional surface for the generalized bathtub model. In Section 3, we derive equivalent formulations with different combinations of time, space, network travel distance, characteristic trip distance, and the cumulative number of trips as coordinates and dependent variables. In Section 4, we present the equivalent formulations for the special cases of Vickrey's and basic bathtub models. In Section 5, we present numerical solution methods and examples based on the equivalent formulations. In Section 6, we present the equivalent trip-based formulations of the generalized bathtub model when the travel demand is given by discrete pairs of entering times and trip distances. Finally, in Section 7, we conclude the study with future research directions.

\section{A new cumulative number of trips and monotone three-dimensional surface of the generalized bathtub model}

In this section, we order trips by their characteristic distances and define a new cumulative number of trips ahead of a trip with a remaining distance at a time instant. This leads to a monotone three-dimensional surface and new $N$-model formulation of the generalized bathtub model.

\subsection{A new cumulative number of trips and monotone three-dimensional surface}

We denote the entering flow-rate at $t$ by $f(t)$ and the percentage of trips with lengths not greater than $x$ by $\phi(t,x)$. Furthermore, the entering rate of trips with lengths not greater than $x$ can be denoted by $f(t,x)=f(t)\phi(t,x)$. We define the number of trips entering the network before $t$ with lengths not greater than $x$ by $F(t,x)$. Thus,
\begin{align}
F(t,x)&=F(0,x)+\int_0^t f(s,x) ds,
\end{align}
where $F(0,x)$ captures the initial condition in terms of the distribution patterns of trips in the network at $t=0$,
or equivalently
\begin{align}
f(t,x)&=\pd{}{t} F(t,x),
\end{align}
where we assume that $F(t,x)$ is differentiable.
As illustrated in \reff{fig:N-functions}, $F(t,x)$ includes all trips whose entering times and lengths are within the  dark gray rectangular region with diagonal lines or a crosshatch pattern in the $(s,y)$ plane. $F(t)\equiv F(t,\infty)$ include all trips  whose entering times and lengths are within the  dark gray rectangular region and the gray region.
Clearly, $f(t,x)\geq 0$ increases in $x$, and $f(t)\equiv f(t,\infty)$. Thus, $F(t,x)$ is non-decreasing in $t$ and $x$. Assuming that $f(t,x)>0$ and $\pd{}{x} \phi(t,x)>0$, the demand function $F(t,x)$ is a three-dimensional, monotonely increasing surface in both $t$ and $x$.

\bfg\bc
\includegraphics[width=4in]{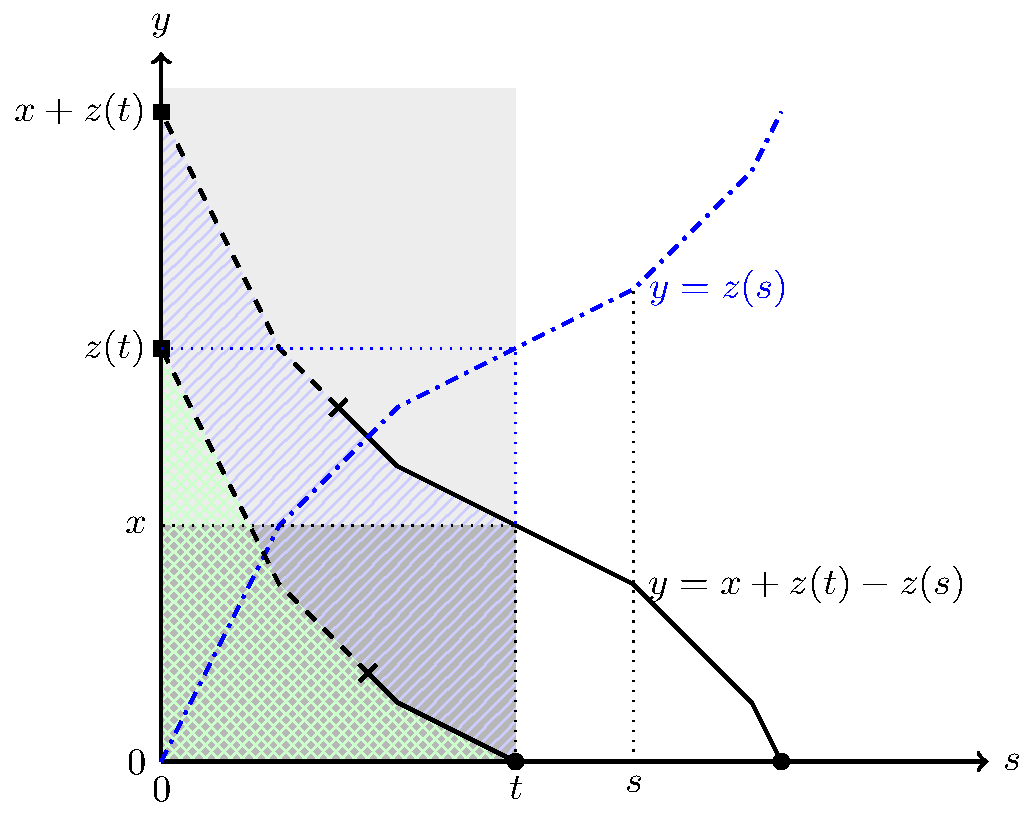}
\caption{Illustration of aggregate variables in the generalized bathtub model}\label{fig:N-functions}
\ec\efg

For the trip flow in a network, we denote the average network travel speed at $t$ by $v(t)$, and the cumulative network travel distance at $t$ by $z(t)$. Assuming that the study period starts at $0$, we have $z(t)=\int_0^t v(s)ds$.

{\bf Assumption 1}. We assume that all trips travel at the same speed at $t$; i.e., $v(t)$.

For trip passing $(t,x)$, its characteristic distance is
\begin{align}
\theta&=x+z(t),
\end{align}
and its remaining distance at $s\leq t$ is
\begin{align}
y&=x+z(t)-z(s).
\end{align}
At time $t$, when comparing two trips, we say one trip is ahead of another if its remaining distance is smaller. Therefore, Trip A being ahead of Trip B is equivalent to Trip A having a shorter characteristic distance. Then the percentage of trips entering the network at $s\leq t$ that are ahead of the trip passing $(t,x)$ is given by $\phi(s, x+z(t)-z(s))$.

We denote the cumulative number of trips entering before $t$ and ahead of the trip at $(t,x)$ by $N(t,x)$. We refer to this as the cumulative flow, as it plays the same role as the cumulative flow for the LWR model \citep{makigami1971traffic}. It is worth noting that the cumulative flow remains constant along a vehicle's trajectory for the LWR model under the FIFO rule. However, this may not be the case for the generalized bathtub model, since the cumulative flow generally increases along a trip trajectory. Here $N(t,x)$ represents a unified, cumulative order for a trip with a remaining distance of $x$ at $t$, since $N(t,x)$ trips would complete before it. 
From the definition we have 
\begin{align}
N(t,x)&= F(0,x+z(t))+\int_0^t f(s, x+z(t)-z(s)) ds, \label{def:N}
\end{align}
As illustrated in \reff{fig:N-functions}, $N(t,x)$ includes all trips whose entering times and lengths are within the diagonally lined gray region and the crosshatched gray region. $N(t,0)$ includes those within the crosshatched gray region.

Then we have the following theorem.
\begin{theorem} \label{thm:N-def}
The cumulative number of trips $N(t,x)$ has the following properties.
\begin{itemize}
\item $N(0,0)=0$,  $N(0,x)=F(0,x)$, and $N(t,\infty)=F(t)$.
\item When $f(t,x)\geq 0$ and $\pd{}{x} \phi(t,x)\geq 0$, $N(t,x)$ is a three-dimensional, monotonically increasing surface in both $t$ and $x$.
\end{itemize}
\end{theorem}

If a trip's remaining distance is less than or equal to $0$, we consider it completed and out of the network. We define the number of completed trips at $t$ by $G(t)$. Then we have
\begin{align}
G(t)&=N(t,0).
\end{align}
We denote the number of active trips at $t$, whose remaining distances are greater than 0, by $\lambda(t)$. Then $\lambda(0)=F(0)$, the total number of trips that have entered the network by $t$ is $F(t)$, and
\begin{align}
\lambda(t)&=F(t)-N(t,0). \label{def:lambda}
\end{align}

\subsection{$N$-model}

\begin{lemma} 
From \refe{def:N}, we have the following partial differential equation:
\bqn
\pd{}t N(t,x)-v(t) \pd{}x N(t,x)&=&f(t,x). \label{N-diff}
\eqn
\end{lemma}

{\bf Assumption 2}. We assume that there exists a network fundamental diagram \citep{godfrey1969mechanism,geroliminis2008eus}, which determines the network travel speed, $v(t)$, from the traffic density per lane, $\lambda(t)/L$, with $L$ as the total lane-miles in the network. Specifically,
\begin{align}
v(t)&=V\left(\frac{\lambda(t)}L\right).\label{NFD}
\end{align}

From \refe{N-diff}, \refe{def:lambda}, and \refe{NFD}, we obtain the following new formulation of the generalized bathtub model:
\bqn
\pd{}t N(t,x)-V\left(\frac{F(t)-N(t,0)}L\right) \pd{}x N(t,x)&=&\pd{}t F(t,x), \label{N-model}
\eqn
where $F(t,x)$ is given as the demand. Hereafter we refer to \refe{N-model} as the $N$-model.

We define $K(t,x)$ as the number of active trips with remaining distances not smaller than $x$ at $t$. Then 
\begin{align}
K(t,x)&=F(t)-N(t,x).\label{K-N-relation}
\end{align}
As illustrated in \reff{fig:N-functions}, $K(t,x)$ includes all trips whose entering times and lengths are within the gray region without diagonal lines or a crosshatch pattern.
The $N$-model, \refe{N-model}, leads to the $K$-model in \citep{jin2020generalized}:
\begin{align}
\pd{}t K(t,x)-V\left(\frac{K(t,0)}L\right) \pd{}x K(t,x)&=\pd{}t [F(t)-F(t,x)]. \label{K-model}
\end{align}
Even though the two formulations are equivalent, $N(t,x)$ has the advantage in that it is monotonely increasing in both $t$ and $x$. This property, along with the inverse function theorem, will lead to more equivalent formulations.

\section{Equivalent formulations} \label{sec:equivalent}
In this section, we first exploit the monotone relations among $t$, $z$, $x$, and $\theta$ and derive equivalent formulations of the generalized bathtub model with different combinations of these variables as coordinates. Then we exploit the monotonicity of $N(t,x)$ and derive equivalent formulations with the cumulative flow as a coordinate. 

\subsection{Equivalent flow models in different coordinates}
In this subsection, we derive $N$- and $K$-models with different combinations of $t$, $z$, $x$, and $\theta$ as coordinates. We refer to them as flow models.

\subsubsection{Network travel distance as a coordinate}

Here we assume that the gridlock state does not occur and $v(t)>0$. Thus $z(t)$ strictly increases in $t$, and we denote the inverse function of $z=z(t)$ by $t=\tau(z)$. Thus, we have $v(\tau(z))=\der{}t z(t)=1/ \der{\tau(z)}{z}$, and $(z,x)$ can be used as new coordinates.

In the $(z,x)$ coordinates with $t=\tau(z)$, the cumulative flow $N(\tau(z),x)$ can be defined as
\begin{align}
N(\tau(z),x)&=F(0,x+z)+\int_0^z  \frac {f(\tau(y),x+z-y)}{v(y)} dy,
\end{align}
and the $N$-model can be re-written as
\begin{align}
\pd{}z N(\tau(z),x)-\pd{}x N(\tau(z),x)&=\pd{}z F(\tau(z),x), \label{N-z-x-model}
\end{align}
where 
\begin{subequations}
\begin{align}
G(\tau(z))&=N(\tau(z),0),\\
\lambda(\tau(z))&=F(\tau(z))-G(\tau(z)),\\
v(\tau(z))&=V(\lambda(\tau(z))/L ).
\end{align}
\end{subequations}

From \refe{K-N-relation}, we can obtain the corresponding $K$-model in the $(z,x)$ coordinates:
\begin{align}
\pd{}z K(\tau(z),x)-\pd{}x K(\tau(z),x)&=\frac 1{v(\tau(z))} \pd{}z [F(\tau(z))-F(\tau(z),x)]. \label{K-z-x-model}
\end{align}

\subsubsection{Characteristic trip distance as a coordinate}
We can see that, among $t$, $x$, and $\theta$, if we know two of them, the other variable is determined. Thus $(t,x)$, $(t,\theta)$, and $(\theta,x)$ can all be valid coordinates. Further, since $t$ can be replaced by $z$ as discussed in the preceding subsection, such that $(z,x)$,  $(z,\theta)$, and $(x,\theta)$ can all be valid coordinates. Among them, the coordinates $(t,x)$ and $(z,x)$  were discussed in Section 2 and the preceding subsection, respectively. Thus, the new coordinates are $(t,\theta)$, $(x,\theta)$, and $(z,\theta)$.

\begin{enumerate}
\item In the $(t,\theta)$ coordinates with $x=\theta-z(t)$, the cumulative flow and the corresponding $N$-model can be written as
\begin{subequations}
\begin{align}
N(t,\theta-z(t))&=F(0,\theta)+ \int_0^t f(s,\theta-z(s))ds,\\
\pd{}t N(t,\theta-z(t))&=f(t,\theta-z(t)) ,\\
G(t)&=N(t,0),\\
\lambda(t)&=F(t)-G(t),\\
v(t)&=V(\lambda(t)/L ).
\end{align}
\end{subequations}

\item In the $(x,\theta)$ coordinates with $x\leq \theta$ and $t=\tau(\theta-x)$, the $N$-model can be written as
\begin{subequations}
\begin{align}
N(\tau(\theta-x),x)&=F(0,\theta)+ \int_0^{\theta-x}  \frac{f(\tau(y),\theta-y)}{v(\tau(y))} dy ,\\
\pd{}x N(\tau(\theta-x),x)&=-\frac{f(\tau(\theta-x),x)}{v(\tau(\theta-x))}, \label{N-x-theta-model}\\
G(\tau(\theta-x))&=N(\tau(\theta-x),0),\\
\la(\tau(\theta-x))&=F(\tau(\theta-x))-G(\tau(\theta-x)),\\
v(\tau(\theta-x))&=V(\lambda(\tau(\theta-x))/L ).
\end{align}
\end{subequations}

\item In the $(z,\theta)$ coordinates with $z\leq \theta$, $t=\tau(z)$, and $x=\theta-z$, we have
\begin{subequations}
\begin{align}
N(\tau(z),\theta-z)&=F(0,\theta)+ \int_0^z f(\tau(y),\theta-y) \frac 1{v(\tau(y))} dy,\\
\pd{}z N(\tau(z),\theta-z)&=\frac{f(\tau(z),\theta-z)}{v(\tau(z))} ,\\
G(\tau(z))&=N(\tau(z),0),\\
\lambda(\tau(z))&=F(\tau(z))-G(\tau(z)),\\
v(\tau(z))&=V(\lambda(\tau(z))/L ).
\end{align}
\end{subequations}

\end{enumerate}

\subsection{Equivalent $X$- and $T$-models with cumulative order as a coordinate}
Since $n=N(t,x)$ is a monotonically increasing 3-D surface in $t$ and $x$, as for the LWR model \citep{makigami1971traffic,jin2016equivalence}, we can rotate the coordinates and represent it by $x=X(t,n)$, which is the remaining distance of the $n$th trip at $t$, and $t=T(n,x)$, which is the time when the $n$th trip's remaining distance is $x$. Mathematically, from the inverse function theorem, the three variables are inverse functions of each other with a fixed common coordinate. Note that since $K(t,x)$ is not monotone in $t$, there cannot be similar formulations of the $K$-model. 

In the $(t,n)$ coordinates with $n\leq F(t)$ and $N(t,X(t,n))=n$, from $n=N(t,X(t,n))$ and the $N$-model, we have the following $X$-model \footnote{The derivation is straightforward, since $N(t,X(t,n))=n$ leads to $\pd{N}t+\pd{N}x \pd{X}t=0$ and $\pd{N}x \pd{X}n=1$, and \refe{N-model} leads to $\pd{N}t-v \pd{N}x =f$.} 
\begin{subequations}
\begin{align}
\pd{}t X(t,n)+f(t,X(t,n)) \pd{}n X(t,n)+v(t)&=0,
\end{align}
where
\begin{align}
G(t)&=X^{-1}(t,0),\\
\lambda(t)&=F(t)-G(t),\\
v(t)&=V(\lambda(t)/L),
\end{align}
\end{subequations}
where $G(t)=N(t,0)=X^{-1}(t,0)$ is the solution of $X(t,N(t,0))=0$. That is, at trip $X^{-1}(t,0)$, the remaining distance is $0$.

Similarly, in the $(n,x)$ coordinates, from $N(T(n,x),x)=n$ and the $N$-model, we have the following $T$-model
\begin{subequations}
\begin{align}
f(T(n,x),x) \pd{}n T(n,x)-v(T(n,x)) \pd{}x T(n,x)&=1,
\end{align}
where
\begin{align}
G(T(n,x))&=N(T(n,x),0),\\
\lambda(T(n,x))&=F(T(n,x))-G(T(n,x)),\\
v(T(n,x))&=V(\lambda(T(n,x))/L).
\end{align}
\end{subequations}

In summary, all equivalent formulations of the generalized bathtub model can be illustrated in \reff{fig:equivalent-formulations}. In the figure, variables with both lower and upper cases can be both dependent or independent variables, and those with only upper cases can be dependent variables, but not independent variables. Each triangle, except the bottom one with thin lines on all sides, represents a possible formulation, in which the two variables connected by a solid line can form the coordinates, and the third variable is the dependent variable. Here $T/t$ and $Z/z$ are totally equivalent. Thus, there can be five formulations for each of the $K$- and $N$-models, three formulations of each of the $X$- and $\Theta$-models, and two formulations of each of the $T$- and $Z$-models. All of these 20 formulations are equivalent.

\bfg\bc
\includegraphics[width=4in]{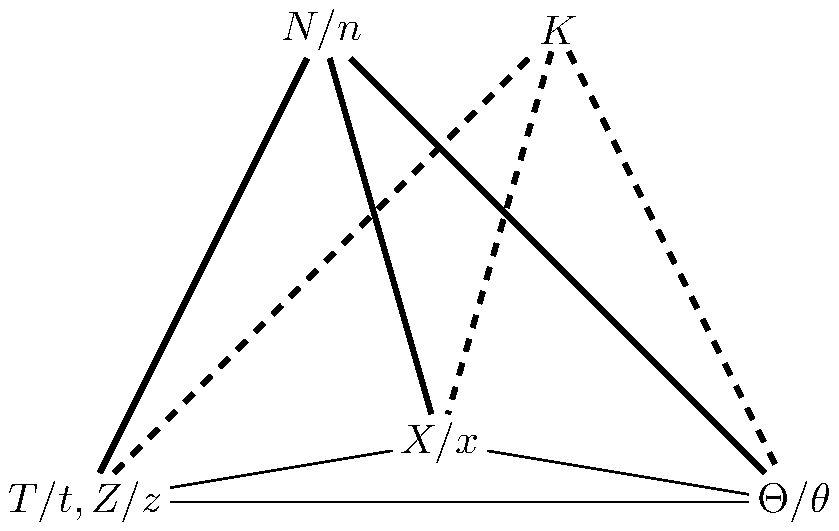}
\caption{Equivalent formulations of the generalized bathtub model}\label{fig:equivalent-formulations}
\ec\efg

\section{Two special cases}
In this section, we present equivalent formulations for two special cases: Vickrey's bathtub model for a time-independent negative exponential distribution of trip distances \citep{vickrey1991congestion,vickrey2020congestion} and the basic bathtub model for a constant trip distance \citep{arnott2016equilibrium,arnott2018solving}.

\subsection{Vickrey's bathtub model: Time-independent negative exponential distribution of trip distances}

When the initial and boundary trips’ distances follow the same time-independent negative exponential distribution with $B$ as the average trip distance, where $F(0,x)=\lambda(0) (1-e^{-\frac xB})$, and $f(t,x)=f(t) (1-e^{-\frac xB})$.

From the definitions of  $\lambda(t)$, $K(t,x)$, and $N(t,x)$, we have
\begin{subequations}
\begin{align}
\lambda(t)&=\lambda(0) e^{-\frac {z(t)}B} + e^{-\frac {z(t)}B} \int_0^t f(s) e^{\frac {z(s)}B} ds,\\
K(t,x)&=\lambda(t) e^{-\frac xB},\\
N(t,x)&=F(t)- \lambda(t) e^{-\frac xB}.
\end{align}
\end{subequations}

Then the $N$-model is equivalent to the following ordinary differential equation
\begin{align}
\der{}t \lambda (t)&=f(t)-\frac 1B  \lambda (t)  V(\lambda (t)/L),
\end{align}
which is the original bathtub model in \citep{vickrey1991congestion,vickrey2020congestion} and independently proposed by other authors without the explicit assumption of negative exponential distribution \citep{small2003hypercongestion,daganzo2007gridlock}.

\subsection{Basic bathtub model: Constant trip distances}
When the initial trips’ distances are not longer than  $\tilde B$  and the entering trips’ distances are constant at  $\tilde B$, the generalized bathtub model is the basic bathtub model considered in \citep{arnott2016equilibrium,arnott2018solving}. In this case, $F(0,x)=F(0)\phi(0,x)$, where $\phi(0,x)=1$ for $x\geq \tilde B$, and 
\begin{align}
\phi(t,x)&\equiv H(x-\tilde B)\equiv \begin{cases} 0, & x<\tilde B; \\1, &x\geq \tilde B,\end{cases}
\end{align}
where $H(\cdot)$ is the Heaviside function. The $N$-model is the same as that  derived in \citep[][Section 5.2]{jin2020generalized}:
\begin{subequations}
\begin{align}
\pd {}t N(t,x)-v(t)\pd{} x N(t,x)&=0,
\end{align}
where the initial $N(0,x)$ is given for $x\in[0,\tilde B]$, and
\begin{align}
N(t,x)&=\cas{{ll} N(0,x+z(t)), & x+z(t) \leq \tilde B; \\ \la(0)+F(\tau(x+z(t)-\tilde B)) , & x+z(t) >\tilde B,}\\
G(t)&=\cas{{ll} N(0,z(t)), & z(t) \leq \tilde B; \\ \la(0)+F(\tau(z(t)-\tilde B)) , & z(t) >\tilde B,}\\
\lambda(t)&=\lambda(0)+F(t)-G(t),\\
v(t)&=V(\lambda(t)/L).
\end{align}
\end{subequations}

It is important to note that trips follow the first-in-first-out principle in this case. All other equivalent formulations can be simplified from those in Section 3 and are omitted here.

Note that, for a single-lane road segment of a length $\tilde B$, if vehicles travel in the decreasing direction of $x\in[0,\tilde B]$, we define $N(t,x)$ as the number of vehicles ahead of $x$, including those that have already exited the road segment. Then the LWR model \citep{lighthill1955lwr,richards1956lwr} can be re-written as:
\begin{subequations}
\begin{align}
\pd {}t N(t,x)-v(t,x)\pd{} x N(t,x)&=0,
\end{align}
where $\pd{} x N(t,x)$ is the traffic density at $(t,x)$, and 
\begin{align}
v(t,x)&=V\left(\pd{} x N(t,x) \right).
\end{align}
\end{subequations}
Here the LWR model is a local Hamilton-Jacobi equation \citep{leclercq2007lagrangian}, but the generalized bathtub model is a nonlocal transport equation.

In the $(z,x)$ coordinates, the $N$-model can be written as
\begin{subequations}
\begin{align}
\pd {}z N(\tau(z),x)- \pd{} xN(\tau(z),x)&=0,
\end{align}
where the initial $N(0,x)$ is given for $x\in[0,\tilde B]$, and
\begin{align}
N(\tau(z),x)&=\cas{{ll} N(0,x+z), & x+z \leq \tilde B; \\ \la(0)+F(\tau(x+z-\tilde B)) , & x+z >\tilde B,}\\
G(\tau(z))&=\cas{{ll} N(0,z), & z \leq \tilde B; \\ \la(0)+F(\tau(z-\tilde B)) , & z >\tilde B,}\\
\lambda(\tau(z))&=\lambda(0)+F(\tau(z))-G(\tau(z)),\\
v(\tau(z))&=V(\lambda(\tau(z))/L).
\end{align}
\end{subequations}

\section{Numerical method and example}\label{sec:numerical}

In this section, we present a numerical method and an example to solve the generalized bathtub model in the $(z,x)$ coordinates.

\subsection{Numerical method in the $(z,x)$ coordinates}

As illustrated in \reff{fig:discretization-method}(a), with regular discretization in the $(z,x)$ coordinates, we set $\Delta x=\Delta z$ as the step-size in $x$ and $z$, and the discrete version of \refe{N-z-x-model} can be written as
\begin{align}
N(\tau(z+\Delta z),x)&=N(\tau(z),x+\Delta x)+f(\tau(z),x) \frac{\Delta z}{v(\tau(z))}. \label{discrete-N-z-x-model}
\end{align}
We divide the range of trip distances $[0,X]$ into $I$ intervals with $\Delta x = \frac X {I}$, where $X$ is the maximum trip distance. We divide the network travel distances to $J$ intervals with $z_j=j\Delta z$ ($j=0,1,\cdots$). Accordingly, we discretize the study period to $J$ time steps and denote them by  $t_j$ ($j=0,1,\cdots$), where  $t_j=\tau(z_j)$. Further the corresponding speed is denoted by $v_j=v(t_j)$. If we denote the time step-size $\Delta t_j= \frac {\Delta z}  {v_j}$, then $t_{j+1}=t_j+{\Delta t}_j$ and $t_{j+1/2}=t_j+\frac 12 {\Delta t}_j$.

\bfg
\bc
\includegraphics[width=6in]{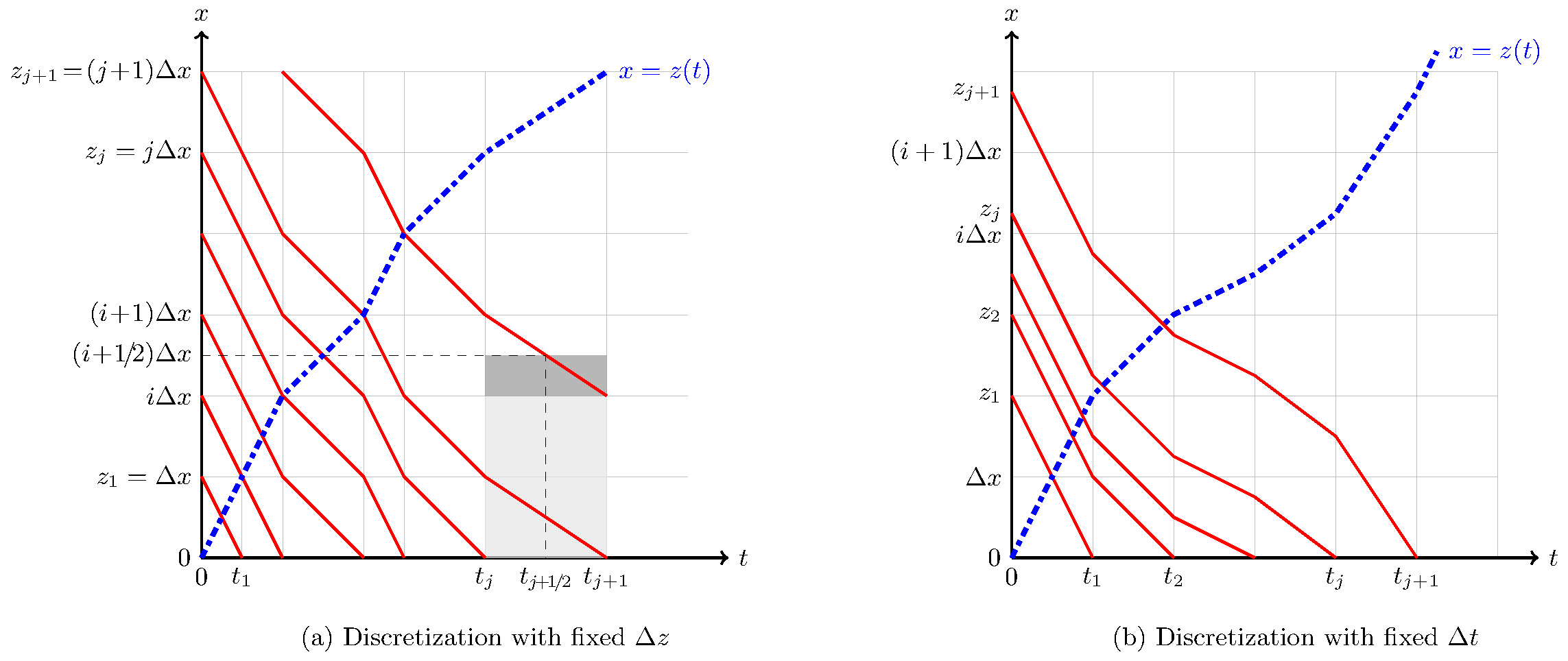}
\ec
\caption{Two discretization methods}\label{fig:discretization-method}
\efg

At $t_j$,  $F(t_j)$ is denoted by $F_j$, $f(t_{j})$ by $f_{j}$, $f(t_{j+1/2})$ by $f_{j+1/2}$, $\phi(t_{j},i\Delta x )$ by $\phi_{j}^i$,  $\phi(t_{j+1/2},(i+1/2)\Delta x )$ by $\phi_{j+1/2}^{i+1/2}$,$N(t_j,i\Delta x )$ by $N_j^i$ ($i=0,\cdots,I$),  the number of completed trips is $G_j=N_0^j$,  and the number of active trips in the network is $\lambda_j=F_j^0 -G_j$. 
From \refe{def:N}, we obtain
\begin{align}
N_{j+1}^i&= N_j^{i+1}+\int_{t_j}^{t_{j+1}} f(s) \phi(s, i\Delta x+v_j (t_{j+1}-s)) ds. \label{def:N-discrete}
\end{align}

Given $N_0^i=F_0 \phi_0^i$, $f_j$, $f_{j+1/2}$, $\phi_{j}^i$, and $\phi_{j+1/2}^{i+1/2}$, we can solve the model from $t_j$ to $t_{j+1}$ with the  first method:
\begin{subequations}\label{method1}
\begin{align}
\lambda_j&=F_j -N_j^0,\\
v_j&=V(\lambda_j/L),\\
\Delta t_j&= \frac {\Delta z}  {v_j},\\
t_{j+1}&=t_j+{\Delta t}_j,\\
F_{j+1}&=F_j+f_{j}  \Delta t_j,\\
N_{j+1}^i&=N_j^{i+1}+f_{j} \phi_{j}^{i} \Delta t_j.
\end{align}
\end{subequations}
or the second method:
\begin{subequations}\label{method2}
\begin{align}
\lambda_j&=F_j -N_j^0,\\
v_j&=V(\lambda_j/L),\\
\Delta t_j&= \frac {\Delta z}  {v_j},\\
t_{j+1}&=t_j+{\Delta t}_j,\\
t_{j+1/2}&=t_j+\frac 12 {\Delta t}_j,\\
F_{j+1}&=F_j+f_{j+1/2}  \Delta t_j,\\
N_{j+1}^i&=N_j^{i+1}+f_{j+1/2} \phi_{j+1/2}^{i+1/2} \Delta t_j.
\end{align}
\end{subequations}

The first method described in \refe{method1} is equivalent to the differential form in \citep{jin2020generalized}. In this method, from \reff{fig:discretization-method}(a) we can see that, if $f(t)$ and $\phi(t,x)$ are constant between $t_j$ and $t_{j+1}$, then $N_{j+1}^i$ equals $N_j^{i+1}$ plus the integration of $f(t)\phi(t,x)$ in the gray rectangle. In contrast, the second method in \refe{method2} adds the integration of $f(t)\phi(t,x)$ in the darker, small rectangle. Comparing the two methods, we can see that the first method systematically under-estimates  $N_{j+1}^{i}$, which leads to over-estimated $\lambda_{j+1}$. Therefore, the network speeds solved by the first method are lower than the theoretical solutions, and this method can lead to artificial gridlock, in which the network travel speed becomes zero due to the under-estimated cumulative flows.

\reff{fig:discretization-method}(b) illustrates regular discretization in the $(t,x)$ coordinates with fixed time and space step sizes, and a corresponding numerical method can be devised, but is omitted here for brevity.

\subsection{An example}
In this subsection, we use the aforementioned numerical method to solve the same example as in \citep{jin2020generalized}. The network has $L=10$ lane-miles and the following speed-density relation: 
\begin{align*}
V(\rho)&=\min\{30, \frac{750} \rho, 10(\frac{200} \rho -1) \}.
\end{align*}
 Initially, the network is empty. The in-flux during a peak period has a trapezoidal shape:
\begin{align*}
f(t)&=\max\{0, \min\{ 10000 t, 4000, 10000(1-t) \}\},
\end{align*}
and the entering trips' distances follow a time-dependent uniform distribution:
\begin{align*}
\phi(t,x)&=\min\{1, \frac x{2 \tilde B(t)} \},
\end{align*}
where $\tilde B(t)$ is the average trip distance at time $t$:
\begin{align*}
\tilde B(t)&=2+\max\{0, \min\{ 7.5 t, 3, 7.5(1-t) \}\}.
\end{align*}
Thus, the maximum trip distance is $X=10$ miles, and both the in-flux and average trip distance are symmetric in time and reach their respective maximum values for $t\in [0.4, 0.6]$ hours. 

\begin{figure}[htbp]
    \centering
    \begin{subfigure}[b]{0.48\textwidth}
        \centering
        \includegraphics[width=\textwidth]{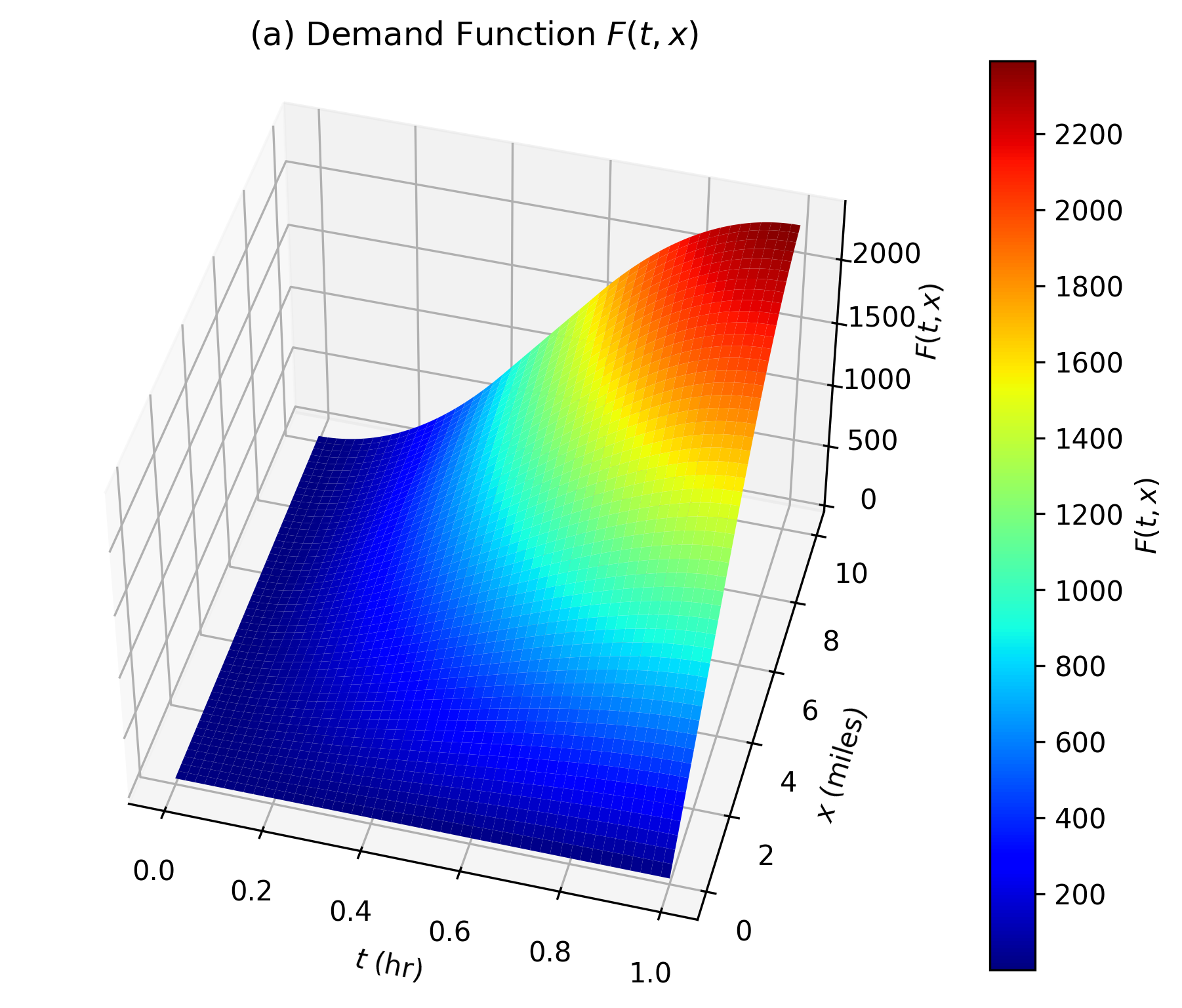}
        \label{fig:f-tx-surface}
    \end{subfigure}
    \hfill
    \begin{subfigure}[b]{0.48\textwidth}
        \centering
        \includegraphics[width=\textwidth]{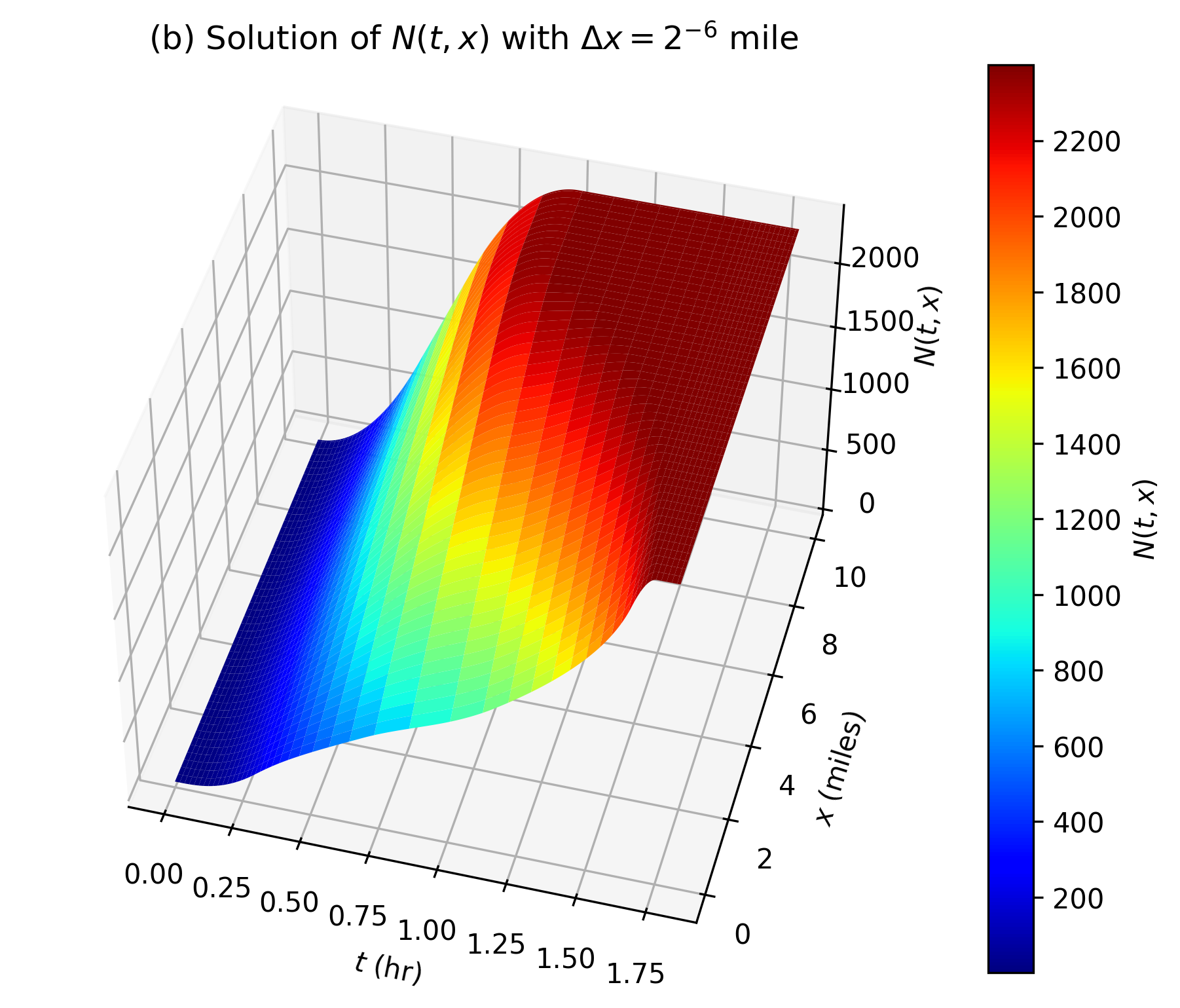}
        \label{fig:gbm-surface}
    \end{subfigure}
    \caption{Monotone three-dimensional surfaces of $F(t,x)$ and $N(t,x)$}
    \label{fig:combined-surfaces}
\end{figure}

The three-dimensional surface $F(t,x)$ is shown in \reff{fig:combined-surfaces}(a). With $\Delta x =2^{-6}$ mile and $I=640$, we solve the generalized bathtub model with the second method and obtain the three-dimensional surface in \reff{fig:combined-surfaces}(b). As shown in the figures, both surfaces are monotonically increasing in both $t$ and $x$, consistent with the analytical results in Section 2.

\begin{figure}[htbp]
    \centering
\includegraphics[width=\textwidth]{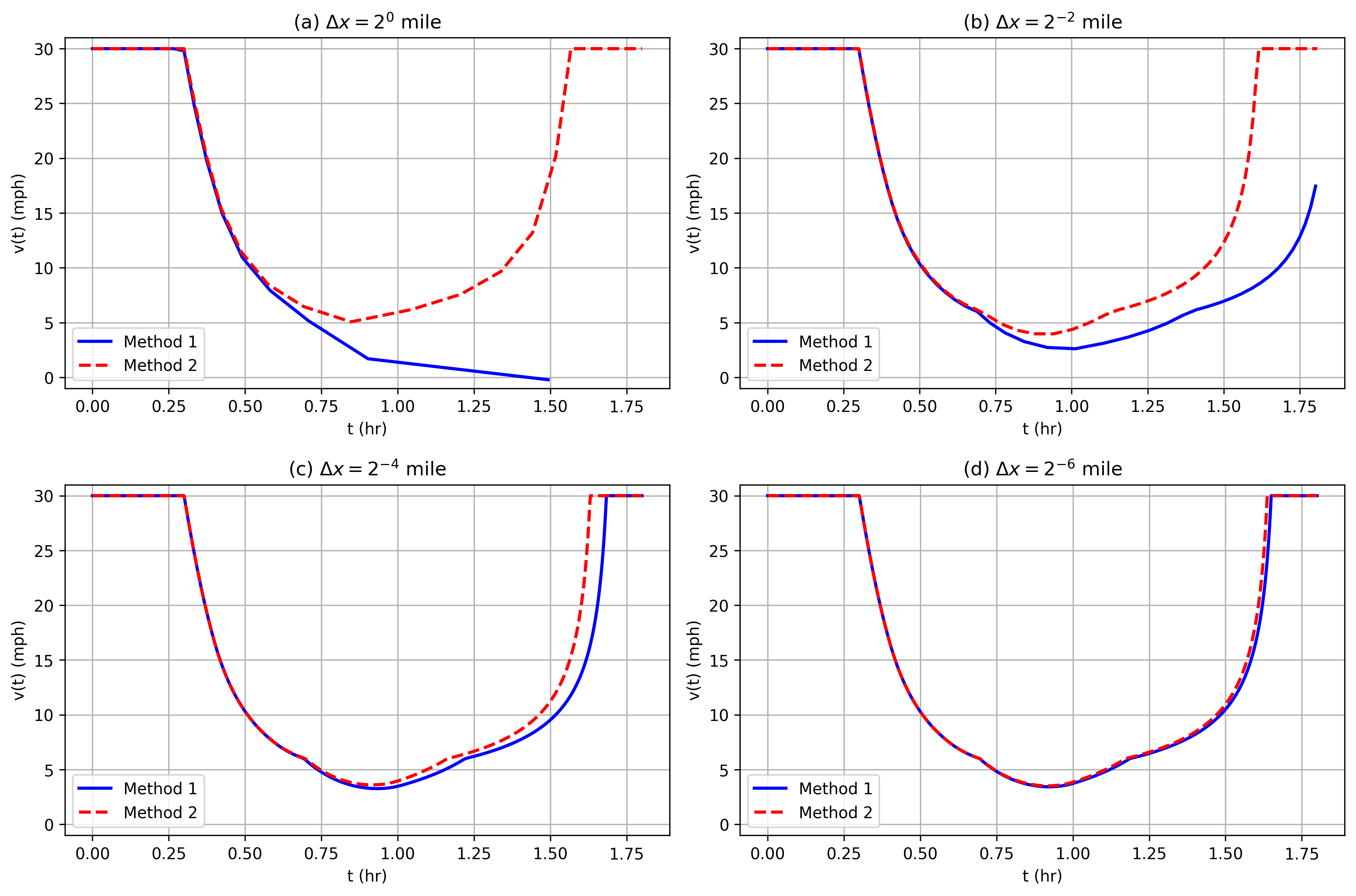}
\caption{Comparison of two numerical methods}
    \label{fig:cGBM_N_z_x_comparison}
\end{figure}

With different space step-size, $\Delta x$, the solutions of $v(t)$ by the two methods are shown in \reff{fig:cGBM_N_z_x_comparison}. The figures confirm that the first method systematically under-estimate the network travel speeds. When $\Delta x=1$ mile, the first method leads to artificial gridlock at $t=1.5$ hr, after which the numerical method fails to yield meaningful result. In contrast, the second method is more accurate and avoids such artificial gridlock solution for this example. 

\begin{figure}[htbp]
    \centering
    \begin{subfigure}[b]{0.48\textwidth}
        \centering
        \includegraphics[width=\textwidth]{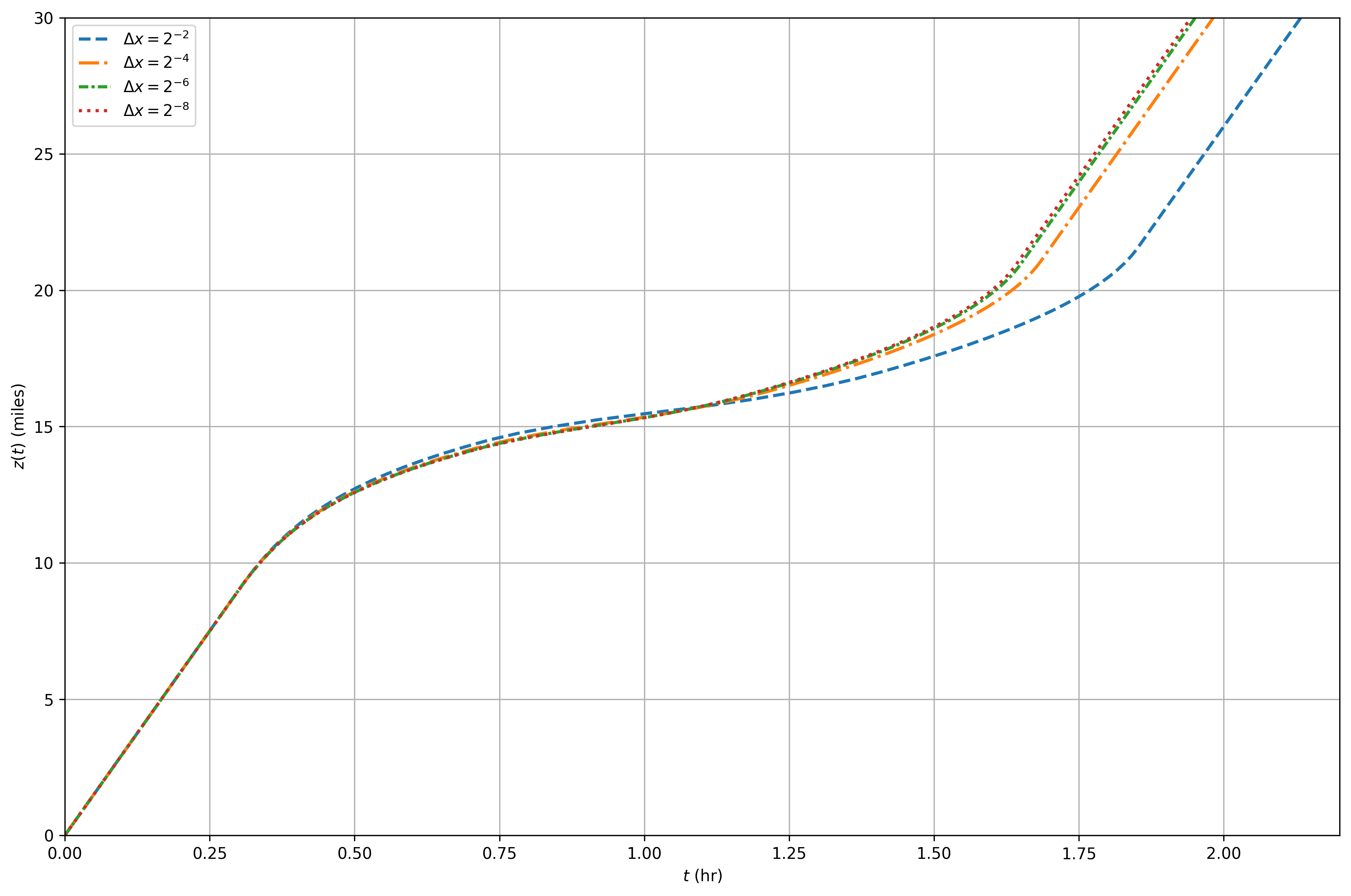}
        \caption{The first method}
    \end{subfigure}
    \hfill
    \begin{subfigure}[b]{0.48\textwidth}
        \centering
        \includegraphics[width=\textwidth]{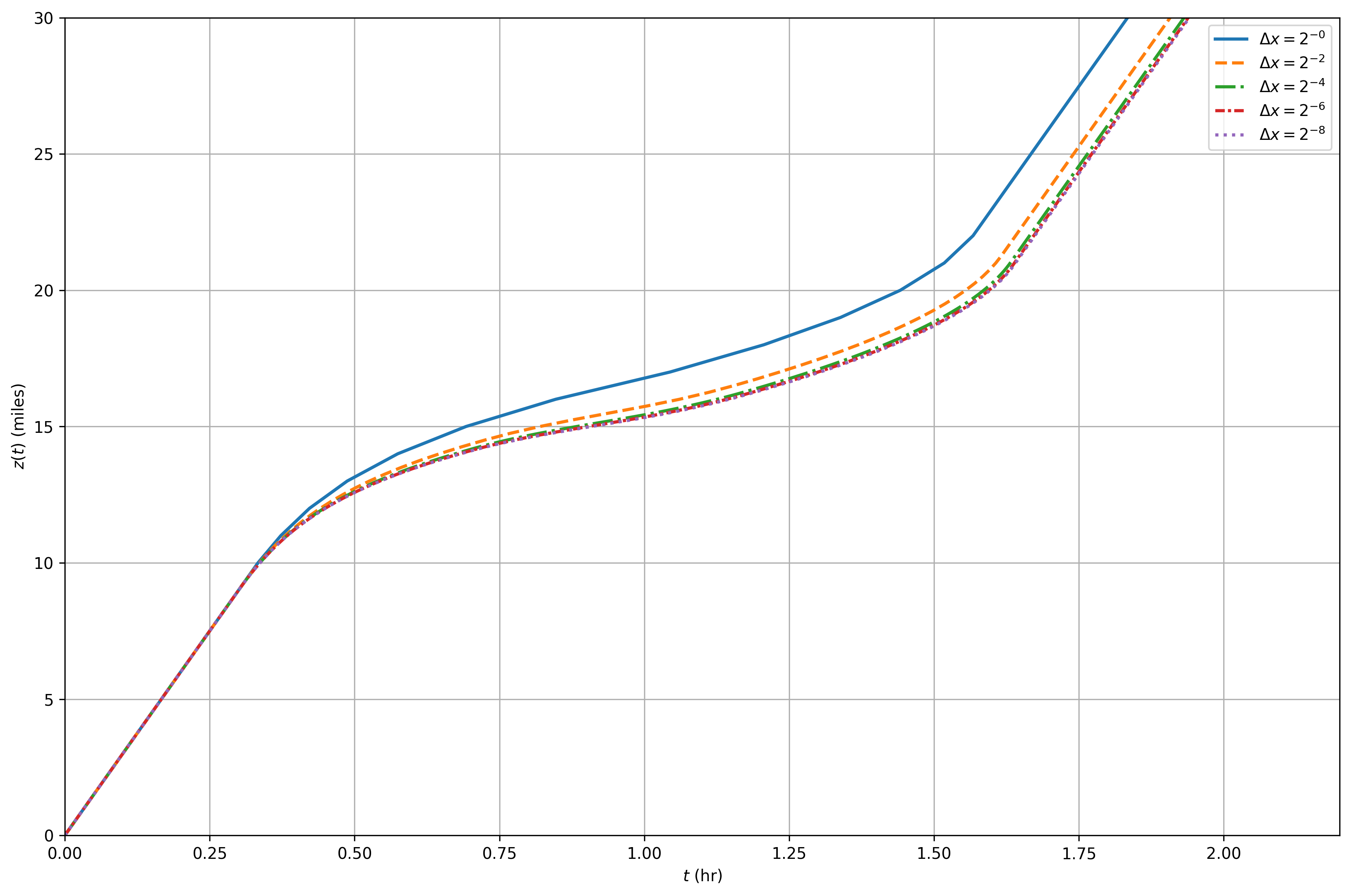}
        \caption{The second method}
    \end{subfigure}
    \caption{Solutions of $z(t)$ for different $\Delta x$ by two different methods}
    \label{fig:GBM_N_z_x_convergence}
\end{figure}

With different spatial step sizes, $\Delta x$, \reff{fig:GBM_N_z_x_convergence} shows that the solutions of $z(t)$ converge as $\Delta x$ diminishes for both methods, suggesting that the numerical solutions converge to the theoretical solution. By comparing the times when the network travel distance reaches 30 miles, we obtain an approximate convergence rate of 1 for both methods. Furthermore, from \reff{fig:GBM_N_z_x_convergence}(b), we can observe that $z(t)$ decreases with $\Delta x$ at any time $t$ in the second method, suggesting that this method systematically overestimates $v(t)$ and does not create artificial gridlock.

\section{Trip-based formulations}

In the preceding sections, the demand pattern is given by continuous functions in $f(t)$ and $\phi(t,x)$, which lead to a smooth surface $F(t,x)$. In this section, we present the equivalent trip-based formulations of the generalized bathtub model when the travel demand is given as discrete pairs of entering times and distances of individual trips. The cumulative flow $N(t,x)$ and the corresponding formulations of the generalized bathtub model are still well defined. However, both $F(t,x)$ and $N(t,x)$ are a non-decreasing piecewise constant functions with a staircase pattern. In this case, we can obtain equivalent trip-based formulations and corresponding numerical methods of the generalized bathtub model.

\subsection{Generalized bathtub model with discrete demand patterns}

We consider a case when the entering times and distances of trips are discrete in the space-time domain. In particular, we assume $M$ time-distance pairs: $(t_m,x_m)$ with $m=1,\cdots,M$. For the purpose of generality, it is possible that multiple trips share the same entering time and distance; thus, we assume that the number of trips sharing $(t_m,x_m)$ is $F_m$, which can be a non-integer number. We refer to all trips with $(t_m,x_m)$ as trip $m$. A set of trips with discrete entering times and trip distances can be used to approximate continuous demand patterns through numerical discretization.

We still define the number of trips entering the network before $t$ with lengths not greater than $x$ by $F(t,x)$, which can be written as
\begin{align}
F(t,x)&= \sum_{m=1}^M F_m H(t-t_m) H(x-x_m), \label{def:F_trip}
\end{align}
where $H(y)$ is the Heaviside function that equals $1$ when $y\geq 0$ and $0$ when $y<0$. Thus, $F(t,x)$ is a non-decreasing piecewise constant function with a staircase pattern, and 
\begin{align}
F(t)&= \sum_{m=1}^M F_m H(t-t_m). 
\end{align}

In this case, $N(t,x)$ can be defined as
\begin{subequations}
\begin{align}
N(t,x)&= \sum_{m=1}^M F_m H(t-t_m) H(x+z(t)-\theta_m), \label{def:N_trip}
\end{align}
where $\theta_m=x_m+z(t_m)$. Thus $N(t,x)$ is also a non-decreasing piecewise constant function with a staircase pattern.
At $t$, $G(t)$ equals the number of completed trips, whose entering times are before $t$ with an extended trip distance not greater than $z(t)$; i.e.,
\begin{align}
G(t)&=N(t,0)=\sum_{m=1}^M F_m H(t-t_m) H(z(t)-\theta_m).
\end{align}
The number of active trips is given by $\lambda(t)=F(t)-G(t)$, and the network travel distance satisfies the following ordinary differential equation:
\begin{align}
\frac{d}{dt} z(t)&=v(t)=V\left( \frac{F(t)-G(t)}{L} \right).
\end{align}
These equations, along with \refe{def:F_trip}, form the generalized bathtub model with discrete demand patterns, which will be referred to as the discrete generalized bathtub model.
\end{subequations}

As $N(t,x)$ is piecewise constant in $t$ and $x$, the inverse functions $X(t,n)$ and $T(n,x)$ are not well-defined. However, we can still derive equivalent formulations in coordinates of $(z,x)$, $(t,\theta)$, $(z,\theta)$, and $(x,\theta)$ as in Section \ref{sec:equivalent}. The corresponding numerical methods in Section \ref{sec:numerical} can be applied to solve the discrete generalized bathtub model.

\subsection{Equivalent trip-based formulations and numerical methods}

The remaining distance of trip $m$ is given by
\begin{align}
X_m(t)&=\theta_m-z(t). \label{def:X_m}
\end{align}
Here $t$ is defined over the entire study period, and $X_m(t)$ represents the extended trajectory of trip $m$, even before it enters the network and after exiting the network.
We further denote the time for trip $m$'s remaining distance to be $x$ by $T_m(x)$, which is given by
\begin{align}
T_m(x)&=\tau(\theta_m-x). \label{def:T_m}
\end{align}
If we denote the cumulative number of trips entering before $t$ and ahead of trip $m$ by $N_m(t)\equiv N(t,X_m(t))$, then we have
\begin{align}
N_m(t)&=N(t,X_m(t))=N(t,\theta_m-z(t))= \sum_{i=1}^M F_i H(t-t_i) H(\theta_m-\theta_i). \label{def:N_m}
\end{align}
Here $N_m(t)$ is a non-decreasing, piecewise constant function in both $t$ and $\theta_m$; that is, the cumulative order of $m$ is non-increasing, and a trip with a longer extended trip distance is always behind that with shorter extended trip distances. That is, the relative ordering of trips remains unchanged. Similarly, we can define $N_m(x)=N(T_m(x),x)$ as the cumulative number of trips that enter before $T_m(x)$ and ahead of trip $m$. However, since $N_m(t)$ is not a strictly monotone function, we cannot define $T_m(n)$ as an inverse function of $N_m(t)$.

\bfg\bc
\includegraphics[width=4in]{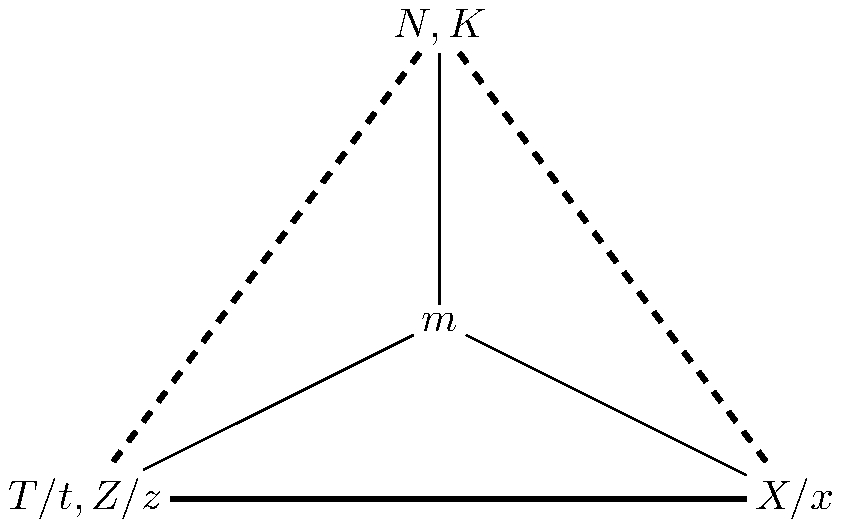}
\caption{Equivalent trip-based formulations of the generalized bathtub model}\label{fig:trip-equivalent-formulations}
\ec\efg

When $v(t)>0$, the network travel distance $z(t)$ can replace time $t$ as an independent variable and the dependent variable. That is, we can define $X_m(\tau(z))$, $Z_m(x)$, and $N_m(\tau(z))$ for trip $m$.
In summary, all equivalent trip-based formulations of the generalized bathtub model can be illustrated in \reff{fig:trip-equivalent-formulations}. In the figure, $m$ serves only as an independent variable, variables with both lower and upper cases can be either dependent or independent variables, and those with only upper cases can be dependent variables, but not independent variables. Then each triangle with at least two solid lines represents a valid trip-based formulation, in which the two lower-case variables can form the coordinates, and the other upper-case variable is the dependent variable.

Furthermore, specific numerical methods can be developed for trip-based formulations of the generalized bathtub model.  \citep{martinez2024priority} developed a priority-queue method for tracking the active trips in a network based on their characteristic distances, and calculated the number of completed trips by comparing the characteristic trip distances against the network travel distance. Effectively, the method keeps track of $N_m(t)$ for all active trips. \citep{pandey2024local} developed an event-based method to track the entering and exiting of individual trips, and update the number of active trips and corresponding network travel speed along with such events. Effectively, the method keeps track of $T_m(0)$ for all active trips.

\section{Conclusion}

In this paper we defined a new cumulative number of trips based on the unchanging ordering of characteristic trip distances; i.e., a trip with a smaller characteristic distance is always ahead with a smaller remaining distance. This new variable leads to the $N$-model and a monotone three-dimensional surface of the generalized bathtub model. This is in contrast to the LWR model on a single lane road, for which vehicles follow the first-in-first-out rule and the cumulative flow naturally leads to a monotone three-dimensional surface. Then based on the monotone relations among various variables, we discussed all possible equivalent formulations of the generalized bathtub model as well as two special cases: Vickrey's and basic bathtub models. We then presented a new numerical method and solved an example to verify that the new cumulative number is monotone in both time and space. We also explained trip-based formulations and corresponding numerical methods when the demand is given as discrete pairs of entering times and distances of individual trips.

The major contributions in this study include the following. First, we defined a cumulative number of trips ahead of a trip with a remaining distance at a time instant and showed that it leads to the new $N$-model and a monotone three-dimensional surface for the generalized bathtub model. This cumulative number determines the relative ordering of trips. Second, based on the monotone relationships among various variables, we used the inverse function theorem to derive 20 equivalent formulations with different coordinates and dependent variables for the generalized bathtub model. This reveals the rich structure and properties of the generalized bathtub model, which could be leveraged for different applications. Third, we devised a new numerical method for solving the generalized bathtub model, which is shown to prevent artificial gridlock caused by numerical errors. Since numerical methods that lead to artificial gridlock overestimate congestion, the new method can prevent the over-estimation of the impacts of various traffic management schemes. Fourth, we presented the equivalent trip-based formulations of the generalized bathtub model when the travel demand is given by discrete pairs of entering times and trip distances.

As equivalent formulations enable simpler analytical and numerical solutions under general and special distribution patterns of trip distances, this study will be helpful for further explorations of the properties, solutions, and applications of the generalized bathtub model in the future. It is well known that, once the generalized bathtub model reaches the gridlock state with $v(t)=0$, the solution afterwards becomes ill-posed, as the network cannot get out of it \citep{daganzo2007gridlock}. In the future, we will examine how to take advantage of various equivalent formulations to develop more accurate and efficient numerical methods so as to avoid gridlock solutions caused by inaccurate numerical methods. We will also be interested in applying various equivalent formulations of the generalized bathtub model to study congestion dynamics, departure time choice, congestion pricing,  traffic control, and planning in multi-modal transportation systems.

\section*{Acknowledgments}
The first author expresses sincere gratitude to the UC ITS Statewide Transportation Research Program (STRP) for their generous financial support. The first author also acknowledges partial support from the National Science Foundation under grant NSF-SCC.CMMI\#1952241, ``SCC-PG: Addressing Unprecedented Community-Centered Transportation
Infrastructure Needs and Policies for the Mobility Revolution''.

\end {document}